\documentclass[10pt]{article}

\usepackage[margin=1.5in]{geometry}
\usepackage{amsmath,amssymb}
\usepackage{graphicx}
\usepackage{booktabs}
\usepackage{cite}
\usepackage[hidelinks]{hyperref}
\usepackage{microtype}
\graphicspath{{./}}

\begin{document}

\title{Simulation-Based Imaging: Learning Acoustic Inverse Problems from Simulated Data}

\author{Luke Bodmer \and E. Bruce Pitman\thanks{\small Department of Materials
Design and Innovation, School of Engineering and Applied Sciences,
University at Buffalo, Buffalo, NY USA}}
\date{}

\maketitle

\begin{abstract}
We introduce Simulation-Based Imaging (SBI), a framework for non-destructive acoustic
imaging in which machine learning models trained entirely on simulated data serve as
real-time solvers for the acoustic inverse problem. A high-fidelity nodal
Discontinuous Galerkin (DG) forward solver generates large training datasets by
randomizing inclusion geometry within a unit-cube domain; a 2D convolutional neural
network then learns a direct mapping from boundary pressure measurements to a
$32\times32\times32$ voxel reconstruction of the interior. The trained model reliably
recovers inclusion position and size from 144 boundary sensors with no prior knowledge
of inclusion count or geometry. Reconstruction error degrades by only 13\% under 5\%
additive measurement noise, and just 17\% of the sensor array (24 of 144 sensors)
suffices for quality within 4\% of full coverage. These results establish SBI as a
viable proof-of-concept imaging device whose complexity resides in software rather than
hardware, opening a path toward cheap, portable, deployable imaging systems.
\end{abstract}


\section{Introduction}
\label{sec:intro}

The ability to image the internal structure of an object without modifying or
destroying it is a technology of enormous practical value: we scan bridges for
structural flaws, the earth for mineral deposits, and human tissue for disease and
injury. Every such technology solves an \emph{inverse problem}: determining the
internal structure that produced a set of observed boundary measurements.

Current solutions face a fundamental constraint: accuracy requires expensive, immobile
hardware. Medical MRI scanners cost \$1--3 million and require purpose-built shielded
rooms \cite{makin2021portable}. Industrial computed-tomography systems for
non-destructive inspection typically cost hundreds of thousands to several million
dollars \cite{castillo2012industry}. MRI scanner density in sub-Saharan Africa averages
0.8 units per million people \cite{anazodo2023framework}, and rural hospitals in the
United States routinely lack advanced imaging infrastructure \cite{radiological2021rural}.

Full Waveform Inversion (FWI) emerged as a physics-driven alternative, iteratively
minimizing a misfit between observed and simulated wavefields to recover material
properties. First proposed for seismic imaging by Lailly~\cite{lailly1983seismic} and
Tarantola~\cite{tarantola1984inversion} in the early 1980s, FWI was later applied to
non-destructive testing~\cite{seidl2018full}, bone
quantification~\cite{bernard2017ultrasonic}, and brain
imaging~\cite{guasch2020full,robins2023dual}. Despite these successes, FWI's iterative
nature is computationally prohibitive: a single 2D brain-slice reconstruction has been
reported to require 32 hours~\cite{guasch2020full}.

Recent work has begun combining deep learning with FWI to reduce computational
cost~\cite{herrmann2023use,zerafa2025synergizing,wang2023machine}, typically using
neural networks as surrogate forward models or post-processing refinement steps.

A parallel line of work trains neural networks as direct solvers for the inverse
problem, bypassing iterative optimization entirely. Fully learned approaches have been
applied across imaging modalities (MRI, CT, seismic, and ultrasound) and have been
extensively reviewed in~\cite{auras2024robustness}. In acoustic non-destructive
testing (NDT), where physical data collection is expensive, there is particular
interest in training on simulated data and deploying on real hardware. McKnight et
al.~\cite{mcknight2024comparison} demonstrated that models trained exclusively on
simulated ultrasonic data initially generalize poorly to real measurements (F1 of
0.394, near chance), but domain adaptation methods including physics-based noise
augmentation recover strong performance (F1 of 0.843), establishing that the
simulation-to-reality gap is closeable. Most closely related to the present work is
Hu et al.~\cite{hu2026deepusct}, who trained an attention-enhanced U-Net on nearly
10,000 simulated wavefields to reconstruct cross-sectional images of industrial
fastener workpieces in near real time, using a modest set of real measurements for
fine-tuning. We extend this paradigm to full 3D reconstruction of arbitrary
inclusion geometries from boundary measurements, using a purely simulation-trained
model with no real-data fine-tuning.

SBI shifts complexity from hardware and runtime computation to an upfront investment
in simulation. Piezoelectric transducers that generate and detect pressure waves cost on the
order of dollars~\cite{piexoelectric2019feng}; the intelligence of the system lives in
a trained model that can run on a laptop or smartphone and produces images in
milliseconds. Once trained, deployment costs essentially nothing. This paper
demonstrates that this approach is viable in a controlled simulation environment and
lays the quantitative foundation for future experimental validation.

\section{Problem Formulation}
\label{sec:problem}

\subsection{Governing Equations}

We model acoustic wave propagation using the linear acoustic system in three spatial
dimensions:
\begin{align}
\frac{\partial u}{\partial t} &= -\frac{1}{\rho}\frac{\partial p}{\partial x},
\label{eq:momentum_x} \\
\frac{\partial v}{\partial t} &= -\frac{1}{\rho}\frac{\partial p}{\partial y},
\label{eq:momentum_y} \\
\frac{\partial w}{\partial t} &= -\frac{1}{\rho}\frac{\partial p}{\partial z},
\label{eq:momentum_z} \\
\frac{\partial p}{\partial t} &= -\kappa\!\left(\frac{\partial u}{\partial x}
  + \frac{\partial v}{\partial y} + \frac{\partial w}{\partial z}\right),
\label{eq:continuity}
\end{align}
where $p$ is acoustic pressure [Pa], $(u,v,w)$ are particle velocity components [m/s],
$\rho$ is mass density [kg/m$^3$], and $\kappa = \rho c^2$ is the bulk modulus [Pa].

\subsection{Imaging Setup}

\begin{figure}[h!]
  \centering
  \includegraphics[width=\textwidth]{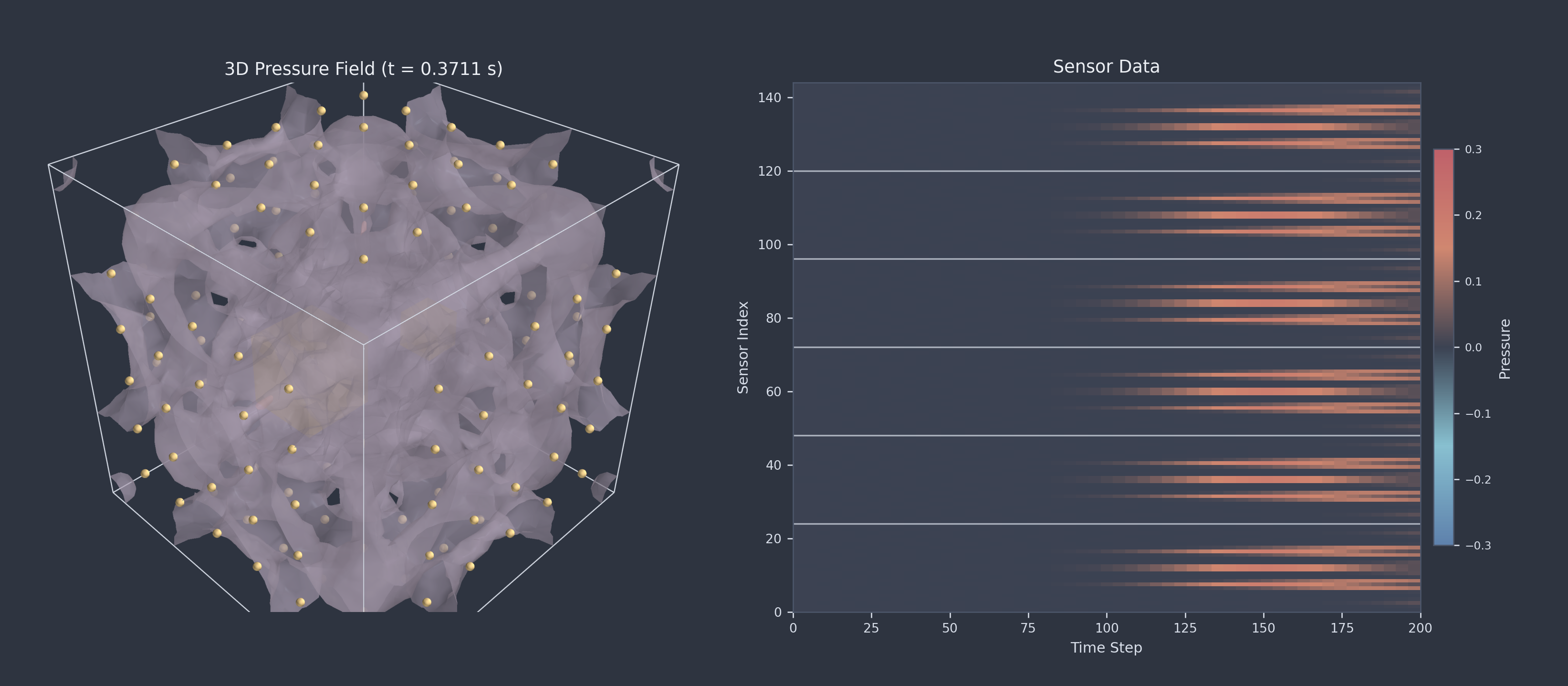}
  \caption{Left: pressure isosurfaces from a DG simulation showing acoustic waves
  propagating through a cubic domain containing an embedded inclusion, with 144 boundary
  sensors (yellow dots on the faces). Right: the resulting 144 sensor time-series that form the
  input to the inverse model.}
  \label{fig:setup}
\end{figure}

The domain is a unit cube $[0,1]^3$ with uniform background material ($\rho_b$, $c_b$).
Inside the cube are one to three smaller cubic inclusions with elevated density and wave
speed ($\rho_i > \rho_b$, $c_i > c_b$). A Gaussian pressure pulse
\begin{equation}
  p_{\mathrm{src}}(t) = A\exp\!\left(-\frac{(t-t_0)^2}{2\sigma^2}\right),
  \quad \sigma = \frac{t_0}{4},
  \label{eq:source}
\end{equation}
is injected from the center of each of the six cube faces in sequence. Pressure
time-series are recorded at 144 sensors distributed in $5\times5$ grids across all six
faces (sensors overlapping the source region are excluded). Fig.~\ref{fig:setup} shows
the domain and sensor arrangement.

The inverse problem is to recover the density field $\rho(\mathbf{x})$ (equivalently,
the position, size, and density of the inclusions) from these 144 time-series alone.
No a priori knowledge of inclusion count, position, or size is assumed.

\section{Forward Model: Nodal Discontinuous Galerkin Solver}
\label{sec:dg}

\subsection{Choice of Discretization}

For 3D wave propagation on domains containing material discontinuities, we require a
method that is geometrically flexible, high-order accurate, and naturally handles
impedance-mismatch interfaces. We implement a nodal Discontinuous Galerkin (DG) finite
element method following Hesthaven and Warburton~\cite{hesthaven2007nodal}. The DG
method uses discontinuous piecewise-polynomial approximations, coupling adjacent
elements through upwind numerical fluxes. Key properties for our application are:

\begin{itemize}
  \item \textbf{Geometric flexibility:} Unstructured tetrahedral meshes conformally
    resolve material interfaces, so density jumps at inclusion boundaries are exact
    rather than smeared.
  \item \textbf{High-order accuracy:} Polynomial order $p$ can be raised without
    remeshing, achieving $\mathcal{O}(h^{p+1})$ convergence.
  \item \textbf{Local conservation:} Each element exactly conserves mass, momentum,
    and energy.
  \item \textbf{Parallelism:} Element-local residual computations are trivially
    parallelizable on GPU hardware.
\end{itemize}

\subsection{Mesh Generation}

Unstructured tetrahedral meshes are generated with Gmsh~\cite{geuzaine2009gmsh}. Each
simulation instance places inclusions at randomized positions; a fresh mesh is generated
per simulation so that element faces align exactly with each inclusion's surface. This
avoids any smearing of the material discontinuity across elements.

\subsection{Reference Element and Basis Functions}

All element-level computations are performed on a reference tetrahedron with vertices
\begin{equation*}
  \mathbf{v}_0{=}(-1,-1,-1),\quad \mathbf{v}_1{=}(1,-1,-1),
\end{equation*}
\begin{equation*}
  \mathbf{v}_2{=}(-1,1,-1),\quad \mathbf{v}_3{=}(-1,-1,1).
\end{equation*}
The affine map to a physical element with vertices $\mathbf{v}_i^{\mathrm{phys}}$ is
\begin{align}
  \mathbf{x}(\xi) = \tfrac{1}{2}\bigl[
    &-(1{+}r{+}s{+}t)\,\mathbf{v}_0^{\mathrm{phys}}
    + (1{+}r)\,\mathbf{v}_1^{\mathrm{phys}} \nonumber\\
    &+ (1{+}s)\,\mathbf{v}_2^{\mathrm{phys}}
    + (1{+}t)\,\mathbf{v}_3^{\mathrm{phys}}\bigr].
  \label{eq:affine_map}
\end{align}

Within each element the solution is represented by $N_p = (p+1)(p+2)(p+3)/6$
high-order nodal basis functions constructed from Jacobi polynomials on the reference
tetrahedron~\cite{hesthaven2007nodal}. Warp-and-blend node positions minimize the
Lebesgue constant, ensuring well-conditioned interpolation. For polynomial order $p=2$
(used in this work) each tetrahedron carries $N_p = 10$ nodes.

Physical-space derivatives are computed using the inverse Jacobian:
\begin{equation}
  \frac{\partial}{\partial x} = \frac{\partial r}{\partial x}\frac{\partial}{\partial r}
    + \frac{\partial s}{\partial x}\frac{\partial}{\partial s}
    + \frac{\partial t}{\partial x}\frac{\partial}{\partial t},
  \label{eq:chain_rule}
\end{equation}
where the metric terms come from inverting the element Jacobian
$\mathbf{J} = \partial(x,y,z)/\partial(r,s,t)$. For straight-sided (affine) tetrahedra
$\mathbf{J}$ is constant within each element, so all precomputed operators
(the Vandermonde matrix $\mathbf{V}$, differentiation matrices $\mathbf{D}_r,\mathbf{D}_s,
\mathbf{D}_t$, and the lift matrix $\mathbf{L}$) need only be formed once on the
reference element and reused for every physical element.

\subsection{Upwind Flux at Material Interfaces}

At element interfaces, the DG method couples discontinuous traces through a numerical
flux. For heterogeneous acoustic media the upwind flux accounts for impedance mismatch
between adjacent materials~\cite{cao2024acoustic}:
\begin{equation}
  \hat{p} = p^{-} + \frac{Z^{+}\bigl(\mathbf{n}\cdot[\mathbf{v}]\bigr) - [p]}
                         {Z^{-} + Z^{+}}\,\kappa^{-},
  \label{eq:flux}
\end{equation}
where $[\cdot]$ denotes the jump across the interface, $Z = \rho c$ is the acoustic
impedance, and superscripts $-$ and $+$ denote interior and exterior traces
respectively. The corresponding velocity flux is obtained symmetrically. This
formulation exactly captures reflections and transmissions at inclusion boundaries
directly from the physics of impedance mismatch.

\subsection{Semi-Discrete System and Time Integration}

The DG discretization of \eqref{eq:momentum_x}--\eqref{eq:continuity} yields a system
of ODEs:
\begin{align}
  \frac{d\mathbf{u}}{dt} &= -\frac{1}{\rho}\nabla_h p - \mathbf{L}(F_u), \label{eq:ode_u}\\
  \frac{dp}{dt}          &= -\kappa\,\nabla_h\cdot\mathbf{u} - \mathbf{L}(F_p), \label{eq:ode_p}
\end{align}
where $\nabla_h$ denotes the discrete gradient/divergence and $\mathbf{L}(F)$ are the
lifted flux contributions from element boundaries.

Time integration uses a low-storage 5-stage 4th-order Runge-Kutta
scheme~\cite{hesthaven2007nodal}. The time step obeys the DG CFL condition
\begin{equation}
  \Delta t \leq \frac{0.9\,h_{\min}}{(2p+1)\,c_{\max}},
  \label{eq:cfl}
\end{equation}
where $h_{\min}$ is the smallest element inscribed diameter and $c_{\max}$ is the
maximum wave speed in the domain.

\subsection{Boundary Conditions}

All six cube faces impose perfectly reflecting (rigid-wall) conditions via
\begin{equation}
  \mathbf{u}^+ = \mathbf{u}^- - 2(\mathbf{n}\cdot\mathbf{u}^-)\mathbf{n}, \qquad p^+ = p^-.
  \label{eq:bc}
\end{equation}
The pressure source \eqref{eq:source} on each face is implemented by overriding the
exterior pressure trace and computing the corresponding velocity adjustment from the
local acoustic impedance.

\subsection{GPU Acceleration and Dataset Generation}

The solver uses CuPy~\cite{okuta2017cupy} as a GPU-accelerated NumPy replacement.
All arrays are transferred to GPU memory before time-stepping; all linear-algebra
operations use GPU-accelerated routines, yielding approximately 10--50$\times$ speedup
over CPU execution.

\begin{figure}[h!]
  \centering
  \includegraphics[width=\textwidth]{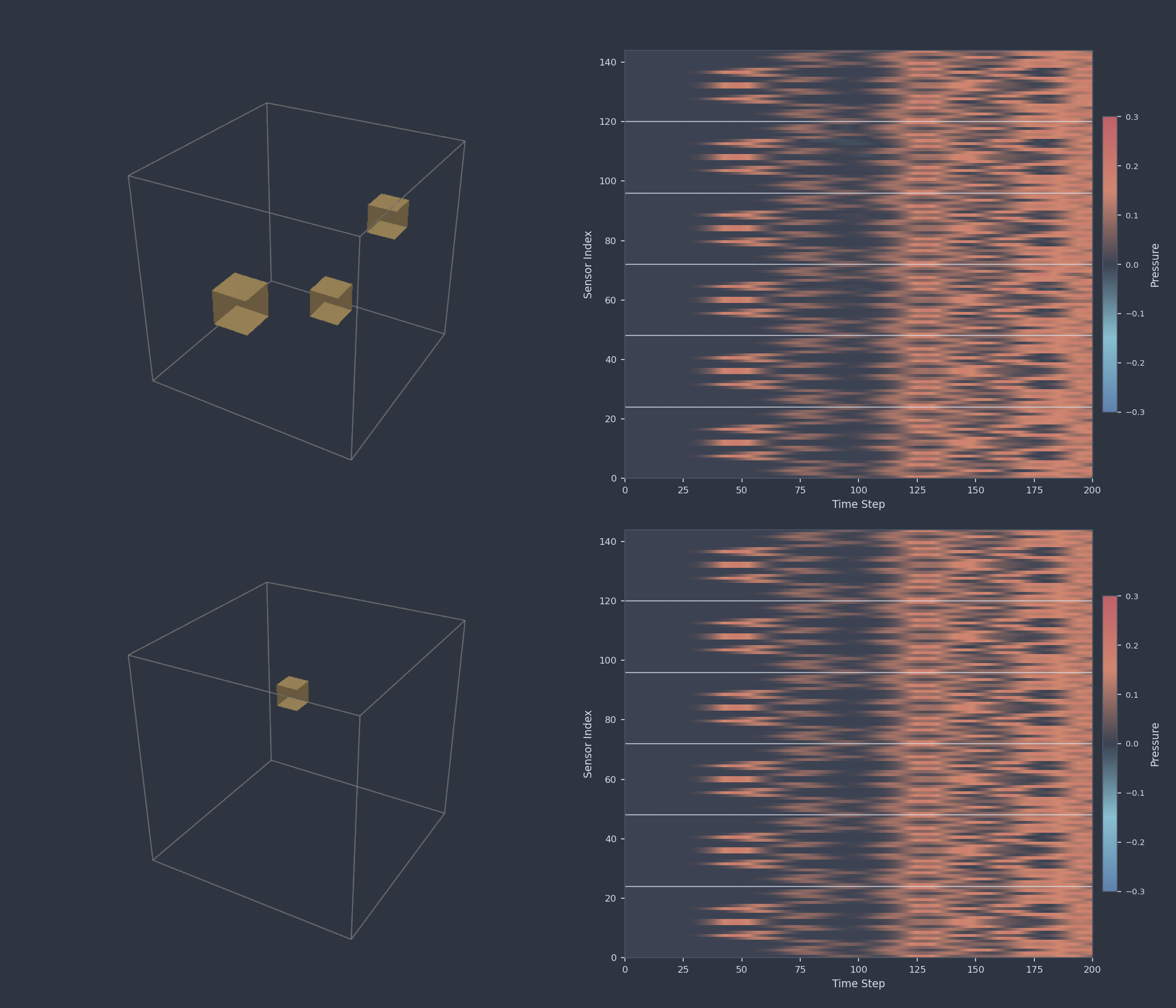}
  \caption{Two representative 3D simulation instances. Each shows the inclusion
  configuration (varying position within the domain) and the corresponding 144
  boundary sensor time-series. The diversity of inclusion positions generates the
  input--output pairs used to train the inverse model.}
  \label{fig:simdata}
\end{figure}

Simulations used $p=2$ with approximately 70,000 tetrahedra, giving $\sim$700,000
scalar degrees of freedom per field. With four field variables (pressure plus three
velocity components) the total is $\sim$2.8 million DOF. A batch of 1000 simulations
ran in parallel on four NVIDIA RTX 3090 GPUs and completed in approximately 48~hours.
Fig.~\ref{fig:simdata} shows a sample of the resulting simulation pairs.

\section{Inverse Model}
\label{sec:inverse}

\subsection{Output Representation}
\label{sec:output}

The inverse model must predict the internal density field $\rho(\mathbf{x})$ from
boundary measurements. We represent the output as a $32\times32\times32$ voxel grid,
then take its 3D discrete Fourier transform to obtain k-space (frequency-domain)
coefficients. The network predicts both real and imaginary k-space components
($2\times32^3 = 65{,}536$ output values); at inference time an inverse FFT recovers
the spatial density image.

This k-space representation offers two practical advantages. First, MSE loss in the
frequency domain naturally weights low-frequency (large-scale) components more heavily
because they have larger magnitude, acting as an implicit regularizer that encourages smooth,
physically plausible reconstructions without requiring a separate penalty. Second,
k-space eliminates constant-valued output columns that arise from buffer regions in the
voxel grid, which cause numerical difficulties for Gaussian Process methods. A direct comparison
confirms that k-space output achieves 3--4\% lower reconstruction error than direct
voxel prediction (Section~\ref{sec:results_output}).

\subsection{Input Preprocessing}
\label{sec:preprocessing}

Raw sensor data consists of 144 time-series of length $\sim$600 timesteps. Three
preprocessing steps reduce dimensionality while preserving the information content
needed for inversion:

\begin{enumerate}
  \item \textbf{Temporal trimming:} The first 41 timesteps are discarded to remove the
    initial source transient before boundary reflections arrive.
  \item \textbf{Temporal downsampling:} Every 4th sample is retained, reducing the
    time axis by $4\times$ while preserving the reflection signatures.
  \item \textbf{Reshape:} The 144 sensor time-series are arranged as a 2D array
    of shape $(\text{sensors}\times\text{timesteps})$ rather than being flattened,
    preserving the spatial layout of the sensor array for 2D convolution.
\end{enumerate}

After preprocessing the input is a $144\times\sim\!154$ array ($\sim$37,000 features
total).

\subsection{Network Architecture}
\label{sec:arch}

The inverse model is a 2D Convolutional Neural Network (2D CNN). Treating the
preprocessed sensor array as a 2D image (sensor index on one axis, time on the
other) lets convolutional filters simultaneously exploit spatial correlations across
sensors and temporal correlations within each sensor's time-series.

The architecture consists of two convolutional blocks followed by adaptive pooling
and a fully connected regressor:

\begin{enumerate}
  \item \textbf{Conv blocks:} Two layers with channel widths [64, 128], kernel size
    $(3\times5)$, stride $(1\times3)$, batch normalization, and ReLU activation.
    Residual (skip) connections are added around each block.
  \item \textbf{Adaptive average pooling:} Output is pooled to a fixed
    $12\times24$ spatial map, yielding $128\times12\times24 = 36{,}864$ features.
  \item \textbf{Regressor:} A single fully-connected layer (512 hidden units, 50\%
    dropout, ReLU) maps to the 65,536-dimensional k-space output.
\end{enumerate}

Table~\ref{tab:arch} summarizes the optimized configuration determined through the
hyperparameter sweeps described in Section~\ref{sec:results_hparam}.

\begin{table}[t]
  \caption{Optimized 2D CNN hyperparameters}
  \label{tab:arch}
  \centering
  \begin{tabular}{ll}
    \toprule
    Parameter & Value \\
    \midrule
    Conv channels      & [64, 128] \\
    Kernel size        & $(3\times5)$ \\
    Stride             & $(1\times3)$ \\
    Pool size          & $(12\times24)$ \\
    Regressor hidden   & 512 \\
    Dropout            & 0.5 \\
    Residual blocks    & Yes \\
    \bottomrule
  \end{tabular}
\end{table}

\subsection{Training Configuration}
\label{sec:training}

The model is trained with AdamW (weight decay $10^{-4}$, initial learning rate
$10^{-4}$) using a OneCycleLR cosine-annealing schedule. MSE loss compares predicted
and true k-space coefficients:
\begin{equation}
  \mathcal{L} = \frac{1}{N}\sum_{i=1}^{N}(k_i - \hat{k}_i)^2,
  \label{eq:loss}
\end{equation}
where $N = 32^3$. Training runs for up to 500 epochs with early stopping (patience 50,
restoring best weights). Batch size is 8; the test fraction is 10\% (100 held-out
simulations). All training was performed on a single NVIDIA RTX 3090 GPU.

\section{Results}
\label{sec:results}

\subsection{Dataset}
\label{sec:dataset}

The training dataset consists of 1000 DG simulations with the following parameters:

\begin{itemize}
  \item Background: $\rho_b = 2.0$~kg/m$^3$, $c_b = 2.0$~m/s.
  \item Inclusions: $\rho_i = 4.0$~kg/m$^3$, $c_i = 4.0$~m/s;
    one to three cubes per simulation.
  \item Inclusion side length: uniform in $[0.1, 0.2]$~m.
  \item Inclusion center: uniform within the interior 90\% of the domain.
  \item Source: Gaussian pulse (frequency 6~Hz) at the center of each face.
  \item Simulation duration: 2~s.
\end{itemize}

The 900/100 train/test split ensures no overlap between training and evaluation
configurations.

\subsection{Qualitative Reconstruction}
\label{sec:results_qual}

\begin{figure}[h!]
  \centering
  \includegraphics[width=0.6\textwidth]{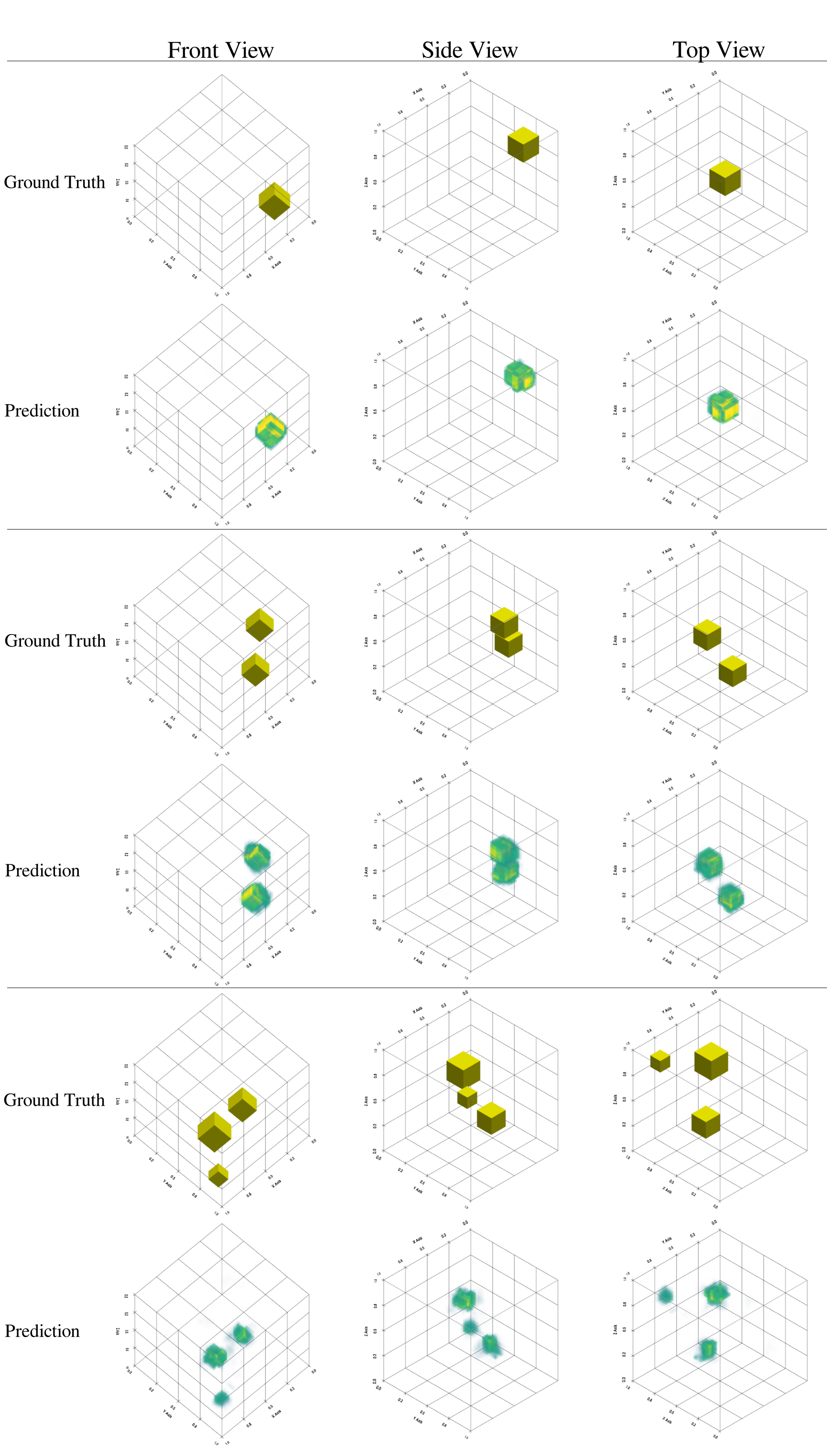}
  \caption{Ground truth (odd rows) vs.\ CNN predictions (even rows) for three held-out
  test examples, viewed from three orthogonal directions (columns: front, side, top).
  The model was trained on 900 simulations with k-space output and a $32^3$ voxel grid.}
  \label{fig:results}
\end{figure}

Fig.~\ref{fig:results} shows three held-out test examples comparing ground-truth
inclusion geometry with CNN predictions. The model successfully localizes inclusions
across the range of positions and sizes in the test set, recovering inclusion location
and approximate extent from 144 boundary time-series alone.

\subsection{Hyperparameter Optimization}
\label{sec:results_hparam}

We swept architectural hyperparameters one at a time, fixing the best value before
varying the next, and measured both MSE test loss and voxel error (sum of
per-voxel absolute reconstruction errors on the test set). Three architecture families
were evaluated: MLP, 1D CNN, and 2D CNN.

The 2D CNN outperformed both alternatives across all hyperparameter settings: the 1D
CNN was 5\% worse and the MLP 8\% worse. The performance gap reflects the information
lost when flattening the sensor array into a 1D sequence: the 2D CNN retains the
spatial layout of the 144 sensors and can exploit correlations between adjacent sensors
that observe the same region of the domain from nearby angles.

Within the 2D CNN sweeps, key findings were:
\begin{itemize}
  \item \textbf{Depth:} A two-layer [64, 128] configuration generalized better than
    deeper networks; additional layers increased overfitting.
  \item \textbf{Pooling:} Asymmetric pooling $(12\times24)$ that preserves temporal
    resolution outperformed symmetric compression.
  \item \textbf{Regressor size:} A compact 512-unit hidden layer outperformed larger
    regressors (1024--4096 units), acting as a bottleneck that forces efficient
    feature compression.
  \item \textbf{Dropout:} Reconstruction error improved monotonically with dropout
    rate; 50\% dropout reduced error by 23\% relative to 10\% dropout.
\end{itemize}

\subsection{Output Representation: K-Space vs.\ Voxel Grid}
\label{sec:results_output}

Table~\ref{tab:kspace} compares direct voxel prediction against k-space prediction
using 5-fold cross-validation. K-space achieves 3\% lower voxel error in 3D despite
MSE loss values that are numerically orders of magnitude larger (due to the large
dynamic range of Fourier coefficients). This apparent paradox arises because MSE on
k-space is dominated by the large DC and low-frequency components, which the network
learns most accurately; the high-frequency residuals are small and contribute little
to spatial reconstruction error.

\begin{table}[t]
  \caption{Output representation comparison (5-fold cross-validation)}
  \label{tab:kspace}
  \centering
  \begin{tabular}{llccc}
    \toprule
    Problem & Repr. & Train Loss & Test Loss & Voxel Error \\
    \midrule
    3D & Voxel   & $5.15{\times}10^{-2}$ & $7.89{\times}10^{-2}$ & $2553 \pm 126$ \\
    3D & K-space & $9.75{\times}10^{2}$  & $1.36{\times}10^{3}$  & $2470 \pm 49$  \\
    \bottomrule
  \end{tabular}
\end{table}

\subsection{Noise Robustness}
\label{sec:results_noise}

To assess behavior under realistic sensing conditions, we trained and evaluated the
optimized 2D CNN at three additive Gaussian noise levels:
\begin{equation}
  X_{\mathrm{noisy}} = X + \mathcal{N}(0,\,(\alpha X_{\mathrm{peak}})^2),
  \label{eq:noise}
\end{equation}
where $\alpha \in \{0\%,5\%,10\%\}$ is the noise fraction relative to peak amplitude.
All experiments used 5-fold cross-validation.

\begin{table}[t]
  \caption{Noise robustness (5-fold cross-validation, mean $\pm$ std)}
  \label{tab:noise}
  \centering
  \begin{tabular}{lccc}
    \toprule
    Noise & Train Loss & Test Loss & Voxel Error \\
    \midrule
    0\%  & $5.03{\times}10^{-2}\!\pm\!1.0{\times}10^{-3}$
         & $8.33{\times}10^{-2}\!\pm\!3.2{\times}10^{-3}$
         & $2469 \pm 50$ \\
    5\%  & $5.21{\times}10^{-2}\!\pm\!1.7{\times}10^{-3}$
         & $9.44{\times}10^{-2}\!\pm\!3.0{\times}10^{-3}$
         & $2783 \pm 114$ \\
    10\% & $5.25{\times}10^{-2}\!\pm\!1.9{\times}10^{-3}$
         & $1.07{\times}10^{-1}\!\pm\!4.5{\times}10^{-3}$
         & $2773 \pm 76$ \\
    \bottomrule
  \end{tabular}
\end{table}

Results (Table~\ref{tab:noise}) show graceful degradation: 5\% noise increases voxel
error by 13\% (2469 to 2783). Notably, 10\% noise produces essentially the same voxel
error as 5\% noise (2773 vs.\ 2783), suggesting the model develops an implicit noise
floor: once noise is above a threshold it neither helps nor significantly hurts
reconstruction quality. Training loss remained nearly constant across noise levels
(5.03 to 5.25$\times10^{-2}$), while test loss increased proportionally with noise,
reflecting the harder inverse problem rather than a failure of optimization. The
small cross-validation standard deviations indicate reproducible, stable performance.

\subsection{Sensor Sparsity}
\label{sec:results_sparsity}

\begin{figure}[h!]
  \centering
  \includegraphics[width=0.6\textwidth]{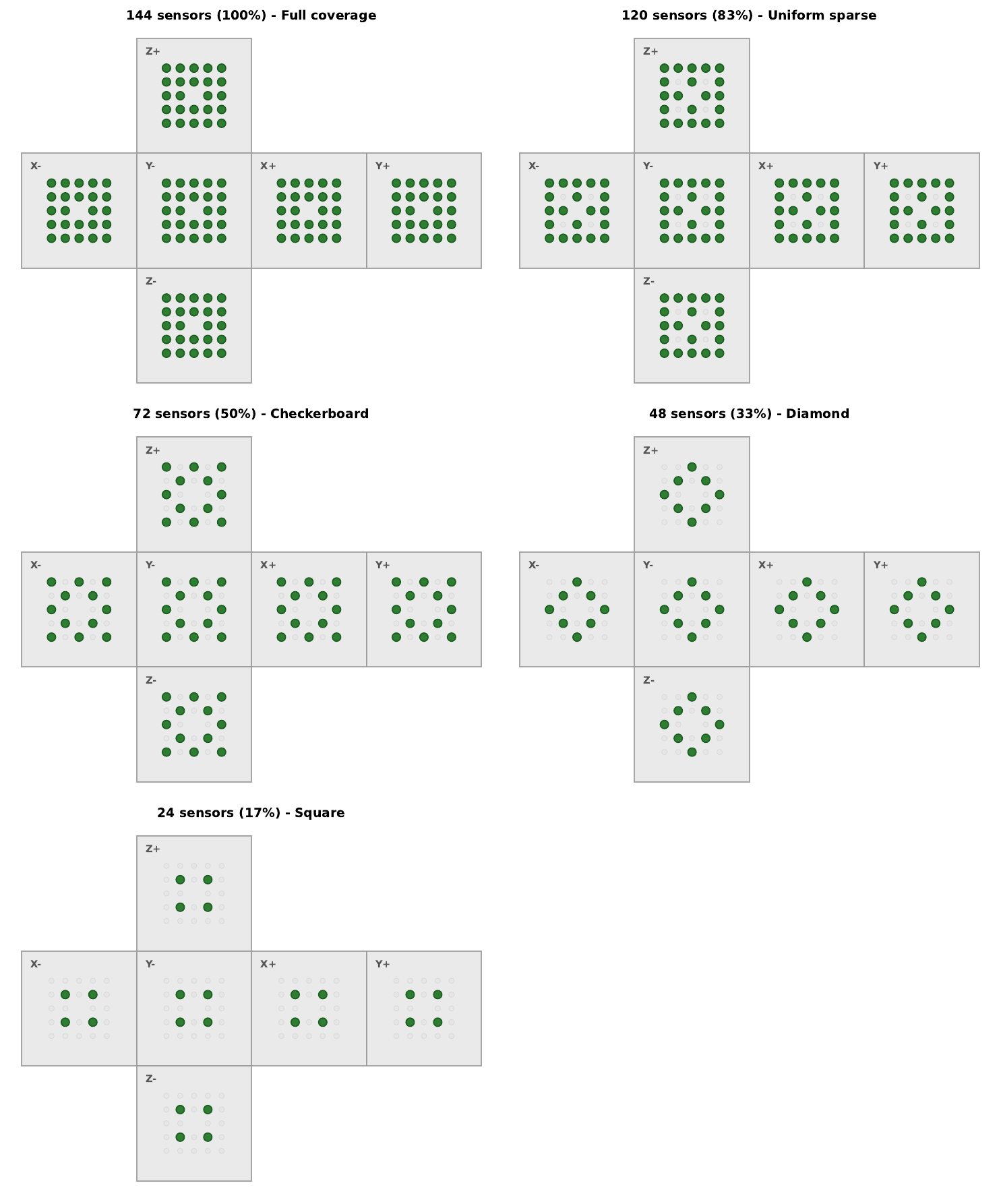}
  \caption{Sensor configurations for the sparsity study, displayed on an unfolded cube.
  Green dots are active sensors; gray dots are inactive. Left to right: full (144),
  uniform-sparse (120), checkerboard (72), diamond (48), square (24).}
  \label{fig:sensors}
\end{figure}

A sensor array of 144 units may be impractical for some deployment scenarios. We
therefore evaluated reconstruction quality as a function of sensor count using five
spatial configurations: full (144), uniform-sparse (120), checkerboard (72), diamond
(48), and square (24). Fig.~\ref{fig:sensors} illustrates each configuration
on an unfolded-cube representation.

\begin{table}[t]
  \caption{Sensor sparsity study (5-fold cross-validation, mean $\pm$ std)}
  \label{tab:sparsity}
  \centering
  \begin{tabular}{llccc}
    \toprule
    Sensors & Pattern & Train Loss & Test Loss & Voxel Error \\
    \midrule
    144 (100\%) & Full            & $5.14{\times}10^{-2}$ & $8.37{\times}10^{-2}$ & $2567 \pm 111$ \\
    120 (83\%)  & Uniform sparse  & $5.31{\times}10^{-2}$ & $8.42{\times}10^{-2}$ & $2686 \pm 67$  \\
    72  (50\%)  & Checkerboard    & $5.21{\times}10^{-2}$ & $8.45{\times}10^{-2}$ & $2564 \pm 110$ \\
    48  (33\%)  & Diamond         & $5.30{\times}10^{-2}$ & $8.34{\times}10^{-2}$ & $2511 \pm 73$  \\
    24  (17\%)  & Square  & $5.32{\times}10^{-2}$ & $8.48{\times}10^{-2}$ & $2544 \pm 115$ \\
    \bottomrule
  \end{tabular}
\end{table}

The results in Table~\ref{tab:sparsity} reveal striking insensitivity to sensor
density. Reconstruction error varies by less than 7\% across all five configurations,
and the 48-sensor diamond pattern actually achieves the \emph{lowest} error, slightly
outperforming full coverage. Reducing from 144 to 24 sensors changes reconstruction
quality by less than 1\% relative to the full-coverage baseline.

This redundancy arises from the physics of the enclosed domain: acoustic energy
injected from each face reflects repeatedly off the rigid walls, so every inclusion is
probed from multiple effective angles regardless of which surface sensors are active.
The CNN learns to extract this redundant information efficiently, making the inverse
model robust to large reductions in hardware sensor count.

\section{Discussion}
\label{sec:discussion}

\subsection{SBI as a Viable Imaging Device}

The results establish three properties that a practical imaging device must have:
(1) it must produce correct images, (2) it must be robust to realistic measurement
imperfections, and (3) it must not require impractical hardware. The CNN reconstructions
in Fig.~\ref{fig:results} satisfy (1); the noise robustness experiments satisfy (2);
and the sensor sparsity results satisfy (3), showing that 17\% of the nominal sensor
count suffices. All three properties hold simultaneously.

The inference time for a trained SBI model is on the order of milliseconds on commodity
GPU hardware, a qualitative shift from the hours required by iterative FWI.

\subsection{Generality of the Approach}

SBI is not specific to linear acoustics or cubic inclusions. Any measurable quantity
that appears in a simulation can be imaged: substituting a viscoelastic model enables
imaging of shear moduli and attenuation; a nonlinear acoustic model (e.g., Westervelt
equation) enables imaging of the nonlinearity coefficient. Substituting a different
sensor geometry or excitation pattern requires only regenerating the training dataset.
The inverse model architecture and training procedure remain unchanged.

Similarly, the DG solver provides the accuracy and geometric flexibility needed to
simulate domains with arbitrarily shaped inclusions, curved boundaries, or multiple
co-existing material regions, all of which are needed for clinical or industrial
applications.

\subsection{Limitations and Future Work}

The central limitation of this work is that all validation uses simulated data. The
simulations employ well-validated numerical methods and should accurately represent
physical wave propagation, but the sim-to-real gap (differences between simulated and
physical sensor responses, material properties, and geometry) must be characterized
experimentally. Physical validation is the most important next step.

Several avenues for improving reconstruction quality within the simulation framework are
also available:

\begin{itemize}
  \item \textbf{Richer source signals:} Multi-frequency pulses or multiple asymmetric
    source positions would probe the domain at multiple length scales and angles,
    reducing inverse-problem ambiguity.
  \item \textbf{Larger training datasets:} The current 1000-simulation dataset was
    constrained by compute budget. More simulations, especially covering a wider range
    of inclusion counts, shapes, and material contrasts, should improve generalization.
  \item \textbf{Advanced architectures:} Transformer attention mechanisms could learn
    which sensors are most informative for each region of the domain. U-Net-style
    encoder-decoder architectures have proven effective for other sensor-to-image tasks
    and warrant evaluation here.
  \item \textbf{Physics-informed training:} Incorporating wave-equation constraints
    into the loss function or network architecture could improve generalization and
    reduce the data requirements.
\end{itemize}

\section{Conclusion}
\label{sec:conclusion}

We have introduced Simulation-Based Imaging (SBI), a framework in which machine
learning models trained on simulated acoustic data serve as real-time solvers for the
3D acoustic inverse problem. Using a nodal DG forward solver on unstructured
tetrahedral meshes to generate 1000 training simulations, and a 2D CNN inverse model
that maps 144 boundary pressure time-series to a $32^3$ density reconstruction, we
demonstrated that:

\begin{itemize}
  \item Inclusion position and size are reliably recovered from boundary measurements
    alone, with no prior knowledge of inclusion count or geometry.
  \item Performance degrades by only 13\% under 5\% measurement noise, and stabilizes
    at 10\% noise.
  \item Only 17\% of the sensor array (24 of 144) is needed for reconstruction quality
    within 4\% of full coverage, enabling large reductions in hardware cost.
\end{itemize}

These results demonstrate that simulation-assisted machine learning models can function
as viable imaging devices: accurate, noise-tolerant, and hardware-efficient. The
complexity of the imaging system lives in software (the trained model and the
simulations that produced it) rather than in specialized hardware. This opens
a path toward cheap, portable, deployable imaging systems for medical diagnostics,
non-destructive testing, and geophysical exploration.

\bibliographystyle{IEEEtran}
\bibliography{references}

\end{document}